\documentclass[final]{myelsarticle}
\usepackage{graphicx}
\newcommand{\beq}{\begin{equation}}
\newcommand{\beqa}{\begin{eqnarray}}
\newcommand{\eeq}{\end{equation}}
\newcommand{\eeqa}{\end{eqnarray}}

\newcommand{\abs}[1]{\vert#1\vert}
\newcommand{\asy}{{\rm asy}}
\newcommand{\ave}{{\rm ave}}
\newcommand{\dd}{{\rm d}}
\newcommand{\e}{{\rm e}}
\newcommand{\eff}{{\rm eff}}

\newcommand{\imax}{{\rm max}}
\newcommand{\imin}{{\rm min}}
\newcommand{\osc}{{\rm osc}}

\begin{document}

\begin{frontmatter}

\title{Revisiting log-periodic oscillations}

\author{Jean-Marc Luck}

\address{Universit\'e Paris-Saclay, CNRS, CEA, Institut de Physique Th\'eorique,
91191~Gif-sur-Yvette, France}

\ead{jean-marc.luck@ipht.fr}

\begin{abstract}
This work is inspired by a recent study of a two-dimensional stochastic fragmentation model.
We show that the configurational entropy of this model
exhibits log-periodic oscillations as a function of the sample size,
by exploiting an exact recursion relation for the numbers of its jammed configurations.
This is seemingly the first statistical-mechanical model where
log-periodic oscillations affect the size dependence of an extensive quantity.
We then propose and investigate in great depth a one-dimensional analogue
of the fragmentation model.
This one-dimensional model possesses a critical point,
separating a strong-coupling phase
where the free energy is super-extensive from a weak-coupling one
where the free energy is extensive and exhibits log-periodic oscillations.
This model is generalized to a family of one-dimensional models with two integer parameters,
which exhibit essentially the same phenomenology.
\end{abstract}

\begin{keyword}
Log-periodic oscillations \sep
Oscillatory critical amplitudes \sep
Fragmentation models \sep
Finite-size scaling \sep
Non-linear recursions \sep
Discrete~scale~invariance
\end{keyword}

\end{frontmatter}

\section{Introduction}
\label{intro}

From the very beginning of the renormalization-group era,
it has been suggested that discrete real-space renormalization-group transformations
might result in the modulation of critical behavior
by a log-periodic function of the distance $t=\abs{T-T_c}/T_c$
to the critical point~\cite{jona,nau,nie}.
Oscillatory critical amplitudes have been later observed and investigated for models
defined on hierarchical lattices~\cite{dee,ddi,dil,bgm,brasov,cg,dg}.
These lattices are self-similar geometric structures
that exhibit discrete scale invariance,
and therefore admit exact renormalization-group transformations.
Similar features are shared by other fractal structures,
on which various models have been shown to exhibit
log-periodic oscillations (see e.g.~\cite{va4,kk,la1,bf5,pm3,ab6,dunne}).
The same phenomenon has also been evidenced as a consequence of the fractal spectra
of some aperiodic structures (see e.g.~\cite{nl25,kt,an00,ca00}).
The first ever mention of log-periodic oscillations seems to date back to 1948,
long before the renormalization group was applied to phase transitions.
In a work devoted to the theory of branching processes,
Harris suggested that iterating a discrete mapping
might yield log-periodic oscillations~\cite{harris}.
These oscillations have then been observed and studied in a more detailed investigation
of a germane combinatorial problem,
namely the enumeration of a class of binary trees~\cite{odlyzko}.

Log-periodic oscillations have since then been reported in a broad variety of situations,
as testified by the overview by Sornette~\cite{s1}.
In many settings including turbulence, fracture, earthquakes, financial crashes,
and quantum gravity,
the occurrence of oscillations is attributed to the emergence
of an approximate discrete scale invariance.
The associated scaling factor is often around two.
Various physical mechanisms, including the period-doubling route to chaos,
have been invoked to explain this phenomenon.
In most of the above situations, the observed amplitude of oscillations is sizeable,
i.e., of the order of 10 percent.
This figure is in stark contrast with the historical example of critical phenomena,
where log-periodic oscillations are typically tiny, with a magnitude of order~$10^{-5}$,
and therefore hard to observe.

The singular behavior of physical quantities
might also be modulated by periodic amplitudes for reasons
that are unrelated to discrete scale invariance.
Confining the discussion to statistical physics,
one-dimensional disordered systems provide a breadth of examples of interest.
There, periodic oscillations are eventually due
to a discrete translation invariance, reflecting
the atomic nature of the underlying lattice~\cite{dh,calan}.
In some models exhibiting anomalous biased diffusion,
the growth law of the mean displacement
is modulated by a log-periodic function of time~\cite{bs2,hk}.
In electron and phonon spectra of disordered chains,
the density of states has exponentially small Lifshitz tails,
whose amplitudes are modulated by periodic functions of $(\Delta E)^{-1/2}$,
where $\Delta E$ is the distance to the band edge~\cite{nl16,nl30}.

A last example, which motivated the present work,
consists of a stochastic fragmentation model introduced in~\cite{bnk}.
A rectangular sample with size $(m,n)$ drawn on the square lattice
is randomly cut into four smaller ones.
The process is then repeated and stops when the system reaches a jammed configuration
where all parts are sticks, i.e., the smaller size of each part equals unity.
Consider for definiteness a square sample of size $n$.
In the first step of the fragmentation process,
the square is cut into four parts whose linear sizes are of the order of~$n/2$.
This observation suggests the emergence of some weak form
of discrete scale invariance with scaling factor two,
that is somehow reminiscent of what is claimed to occur in more complex phenomena
such as turbulence or diffusion-limited aggregation~\cite{s1}.
The main advantage of the above fragmentation model,
which triggered our interest,
is the existence of an exact recursion formula
for the numbers $Z_{m,n}$ of jammed configurations.

The setup of the present paper is as follows.
In Section~\ref{2d}
we revisit the stochastic fragmentation model introduced in~\cite{bnk}.
By exploiting the exact recursion recalled above,
we demonstrate that the configurational entropy based on the numbers $Z_{n,n}$
of jammed configurations on square samples
exhibits tiny log-periodic oscillations in the sample size $n$.
We then introduce and investigate in detail a one-dimensional (1D) toy model
capturing the essential features of the combinatorics of the fragmentation model
(Sections~\ref{1dgal} to~\ref{1dlo}).
In spite of its relative simplicity,
the 1D model turns out to have a rich phenomenology,
with a critical point (Section~\ref{1dcrit})
separating a strong-coupling phase (Section~\ref{1dhi})
from a weak-coupling one (Section~\ref{1dlo}).
In Section~\ref{pers} we consider several other examples of 1D non-linear recursions.
A brief discussion of our findings is given in Section~\ref{disc}.

\section{The fragmentation model revisited}
\label{2d}

\begin{figure}[!ht]
\begin{center}
\includegraphics[angle=0,width=.6\linewidth,clip=true]{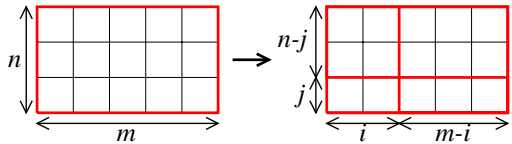}
\caption{\small
First step of the fragmentation of a rectangular sample with $m=5$, $n=3$, $i=2$ and $j=1$.}
\label{frag}
\end{center}
\end{figure}

\begin{figure}[!ht]
\begin{center}
\includegraphics[angle=0,width=.5\linewidth,height=.5\linewidth,clip=true]{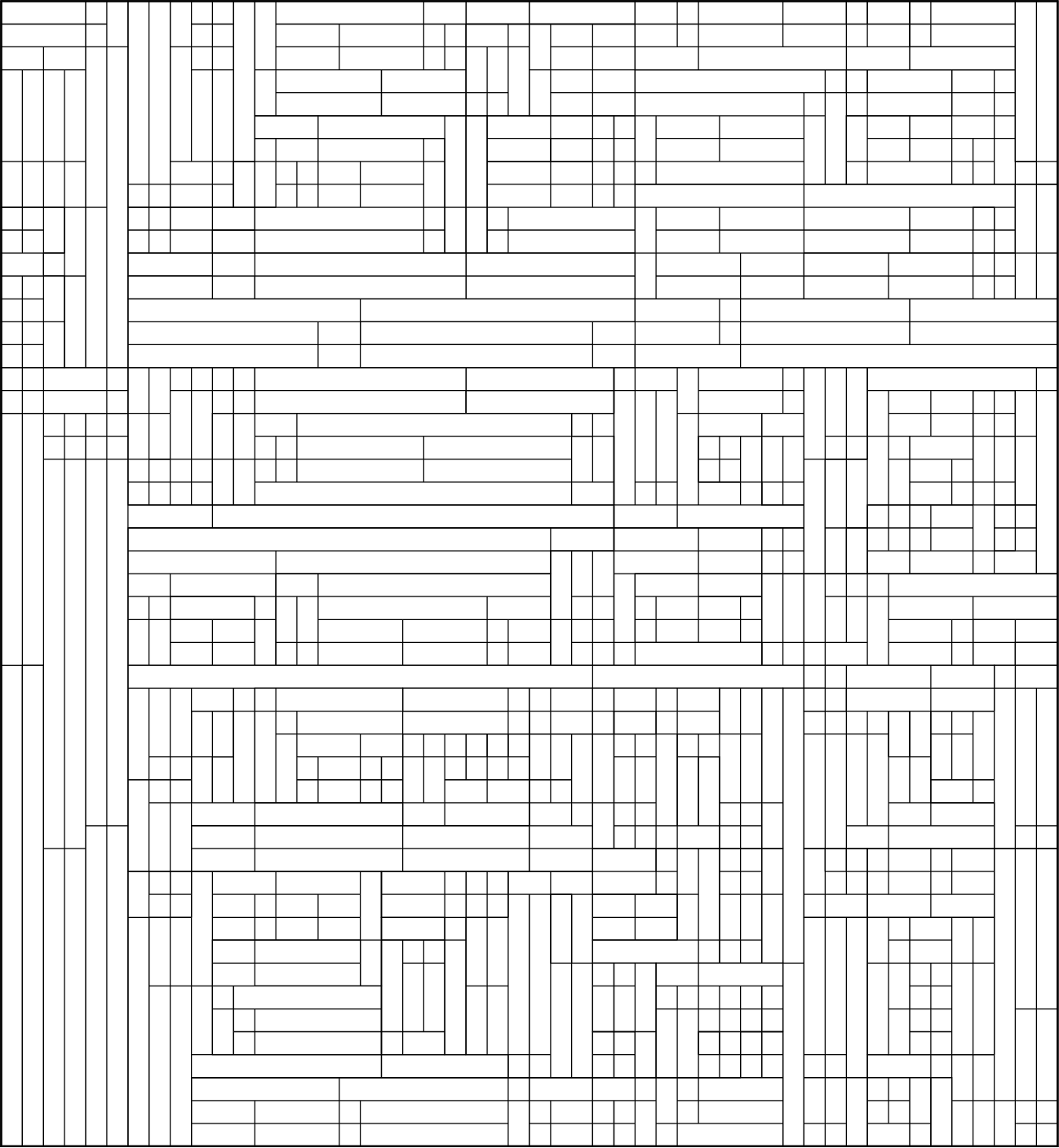}
\caption{\small
A jammed tiling of a square sample of size 50 (courtesy Paul Krapivsky).}
\label{tiling}
\end{center}
\end{figure}

In this section we revisit the second fragmentation model introduced in~\cite{bnk}.
This irreversible stochastic model is defined recursively as follows.
The sample is a rectangle with size $(m,n)$ drawn on the square lattice.
The first step of the fragmentation consists in cutting the rectangle into four smaller ones,
of respective sizes $(i,j)$, $(i,n-j)$, $(m-i,j)$, $(m-i,n-j)$ (see Figure~\ref{frag}).
The integers $i$ and $j$ are chosen uniformly in the ranges $1\le i\le m-1$ and $1\le j\le n-1$.
The process is repeated and stops when the system reaches a jammed configuration
where all parts are sticks, i.e., the smaller size of each part equals unity.
Figure~\ref{tiling} shows a jammed tiling thus obtained on a square of size~50.

This fragmentation model enjoys the property
that the stochastic histories of different rectangles
are mutually independent from the epoch they are formed.
As a consequence, the numbers $Z_{m,n}$ of jammed tiling configurations
on rectangular samples of size $(m,n)$
obey the recursion formula~\cite{bnk}
\beq
Z_{m,n}=\sum_{i=1}^{m-1}\sum_{j=1}^{n-1}Z_{i,j}Z_{i,n-j}Z_{m-i,j}Z_{m-i,n-j},
\label{rec2}
\eeq
with initial conditions $Z_{m,1}=Z_{1,n}=1$ for all $m$ and $n$.
These `initial' values actually express the `final' condition that a jammed tiling is reached
when all parts are sticks, i.e., the smaller size of each part equals unity.
The recursion formula~(\ref{rec2}) determines all the numbers of jammed tilings,
which obey the symmetry $Z_{m,n}=Z_{n,m}$.
The first few of them are listed in Table~\ref{zed}.

\begin{table}[!ht]
\begin{center}
$$
\begin{array}{|c|rrrrrrrr|}
\hline
m\hbox{\large{$\backslash$}} n & 1 & 2 & 3 & 4 & 5 & 6 & 7 & 8 \\
\hline
1 & 1 & 1 & 1 & 1 & 1 & 1 & 1& 1 \\
2 & 1 & 1 & 2 & 3 & 4 & 5 & 6 & 7 \\
3 & 1 & 2 & 4 & 10 & 20 & 36 & 60 & 94 \\
4 & 1 & 3 & 10 & 33 & 98 & 258 & 618 & 1\,379 \\
5 & 1 & 4 & 20 & 98 & 436 & 1\,676 & 5\,848 & 18\,906 \\
6 & 1 & 5 & 36 & 258 & 1\,676 & 9\,524 & 48\,296 & 225\,938 \\
7 & 1 & 6 & 60 & 618 & 5\,848 & 48\,296 & 354\,224 & 2\,387\,112 \\
8 & 1 & 7 & 94 & 1\,379 & 18\,906 & 225\,938 & 2\,387\,112 & 23\,097\,969 \\
\hline
\end{array}
$$
\caption{Numbers $Z_{m,n}$ of jammed tilings on rectangular samples of size $(m,n)$
for $m$ and $n$ up to~8.}
\label{zed}
\end{center}
\end{table}

The numbers $Z_{m,n}$ of tiling configurations grow rapidly with the sample size $(m,n)$.
It is indeed to be expected that their logarithm is extensive,
in the sense that it obeys an area law of the form
\beq
\ln Z_{m,n}\approx S mn,
\label{sarea}
\eeq
where $S$ is the bulk configurational entropy of the model per unit area~\cite{jackle,palmer},
for which the estimate $S\approx0.2805$ is given in~\cite{bnk}.
There are therefore some $10^{304}$ different configurations of jammed tilings
on a square of size 50,
only one of which is shown in Figure~\ref{tiling}.
A more complete thermodynamical expression for $\ln Z_{m,n}$,
including the contributions $S_e$ of the edges of the sample and~$S_c$ of its corners, reads
\beq
\ln Z_{m,n}\approx S mn+2S_e(m+n)+4S_c.
\label{sfull}
\eeq

The asymptotic behavior of $\ln Z_{m,n}$
emerges as a global property of the solution to the recursion~(\ref{rec2}),
that has resisted all our attempts at analysis.
As it turns out, the entropies~$S$, $S_e$ and $S_c$
are modulated by log-periodic oscillations (see below).
Moreover, at variance with the thermodynamical expectation~(\ref{sfull}),
the edge and corner entropies $S_e$ and~$S_c$
have a complex dependence on the aspect ratio $r=m/n$ of the sample (not described here).
None of these peculiar features was mentioned in~\cite{bnk}.

In order to pursue, we have recourse to a numerical solution of the recursion~(\ref{rec2}).
For the sake of definiteness, we focus our attention onto square samples.
As already put forward in Section~\ref{intro},
a square of size~$n$ is cut into four parts whose sizes are of the order of $n/2$
in the first step of the fragmentation process.
This suggests the emergence of some weak form
of discrete scale invariance with scaling factor two,
that is somehow reminiscent of what is claimed to occur in turbulence or in
diffusion-limited aggregation~\cite{s1}.
This intuitive line of reasoning
therefore opens the possibility that the area law~(\ref{sarea})
might be modulated by a 1-periodic oscillatory function of the variable
\beq
x=\frac{\ln n}{\ln 2}.
\label{xdef}
\eeq

Our goal is to demonstrate that log-periodic oscillations are indeed present
in the configurational entropy of the fragmentation model.
For methodological reasons, we begin by considering the ratio
\beq
R_n=\frac{Z_{n,n}}{(Z_{n/2,n/2})^4}
\label{rdef}
\eeq
for even $n$.
The denominator of~(\ref{rdef}) is nothing but the central term
in the expression~(\ref{rec2}) for $Z_{n,n}$, corresponding to $i=j=n/2$.
It is therefore natural to interpret $R_n$ as the effective number of terms
contributing to the recursion~(\ref{rec2}) for $Z_{n,n}$.

\begin{figure}[!ht]
\begin{center}
\includegraphics[angle=0,width=.6\linewidth,clip=true]{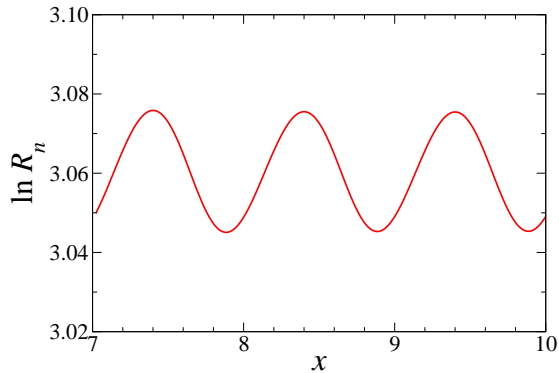}
\caption{\small
Logarithmic plot of the ratio $R_n$ introduced in~(\ref{rdef})
against the logarithmic variable $x$ defined in~(\ref{xdef}).}
\label{rplot}
\end{center}
\end{figure}

Figure~\ref{rplot} shows a logarithmic plot of $R_n$
against the logarithmic variable~$x$ defined in~(\ref{xdef}).
If the numbers $Z_{n,n}$ of tiling configurations were exactly given
by the thermodynamic formula~(\ref{sfull}), we would have
\beq
\ln R_n\approx-4S_en-12S_c.
\eeq
It is obvious from Figure~\ref{rplot} that the edge configurational entropy $S_e$
vanishes for square samples.
Indeed $\ln R_n$ would otherwise exhibit a linear growth in~$n$ with slope $4\abs{S_e}$,
i.e., an exponential growth in $x$.

The quantity $\ln R_n$ plotted in Figure~\ref{rplot} clearly exhibits
1-periodic oscillations in $x$.
Here and throughout the following, we characterize an oscillatory periodic function
$f(x)$ by its average $f_\ave$ over one period and by the relative magnitude of oscillations,
\beq
f_\osc=\frac{f_\imax-f_\imin}{f_\ave}.
\label{oscdef}
\eeq
The data shown in Figure~\ref{rplot} yield
\beq
(\ln R)_\ave\approx3.060,\qquad
(\ln R)_\osc\approx9.8\;10^{-3}.
\eeq
Pre-asymptotic corrections to log-periodic behavior are negligible in the range
shown in Figure~\ref{rplot}.

The corner configurational entropy of a large square can be estimated~as
\beq
S_c=-\frac{(\ln R)_\ave}{12}\approx-0.255,
\label{scres}
\eeq
whereas the effective number $R_n$ of terms entering the recursion~(\ref{rec2})
approaches the rather large but finite limit
\beq
R_\eff=\exp\left((\ln R)_\ave\right)\approx21.32.
\eeq

\begin{figure}[!ht]
\begin{center}
\includegraphics[angle=0,width=.6\linewidth,clip=true]{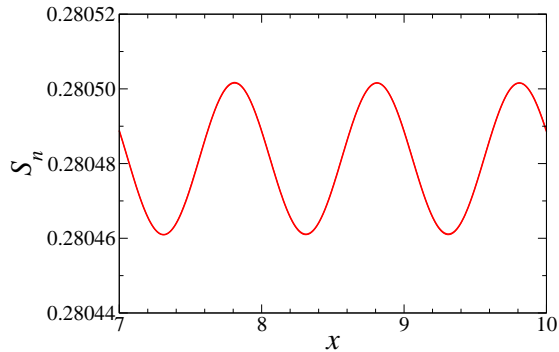}
\caption{\small
Configurational entropy $S_n$ defined in~(\ref{sdef}),
plotted against the logarithmic variable $x$ defined in~(\ref{xdef}).}
\label{splot}
\end{center}
\end{figure}

A more accurate definition of the bulk configurational entropy of square samples is given by
\beq
S_n=\frac{\ln Z_{n,n}-4S_c}{n^2},
\label{sdef}
\eeq
where $S_c$ is taken from~(\ref{scres}).
Figure~\ref{splot} shows that this quantity
exhibits very small but very clear 1-periodic oscillations in the logarithmic variable $x$.
This is seemingly the first instance where
log-periodic oscillations are reported for the size dependence of an extensive quantity
in a statistical-mechanical model.

The subtraction of the corner entropy $4S_c$ has drastically improved convergence,
so that corrections are again negligible in the range shown in Figure~\ref{splot}.
From a quantitative viewpoint, we have
\beq
S_\ave\approx0.280481,\qquad
S_\osc\approx1.44\;10^{-4}.
\eeq
The numerical value of~$S_\ave$ is in full agreement
with the estimate $S\approx0.2805$ given in~\cite{bnk}.
The magnitude $S_\osc$ of oscillations is within the range
commonly observed, e.g.~in the critical properties of models on hierarchical lattices.

\section{1D model: generalities}
\label{1dgal}

In this section we introduce a one-dimensional (1D) analogue
of the combinatorics of jammed configurations
in the fragmentation model investigated in Section~\ref{2d}.
This 1D toy model captures the main features of the fragmentation model,
whereas it is simple enough to be studied in great depth
(see Sections~\ref{1dcrit} to~\ref{1dlo}).
Furthermore, the 1D model has a richer phenomenology
than the original fragmentation model,
with a critical point separating a weak-coupling phase,
where the free energy is extensive and exhibits oscillations,
from a strong-coupling one,
where the free energy is super-extensive and does not manifest oscillations.

The 1D model is defined by the recursion relation
\beq
Z_n=\sum_{k=1}^{n-1}Z_k^2Z_{n-k}^2\qquad(n\ge2).
\label{rec1}
\eeq
This recursion keeps the essential characteristics of its two-dimensional analogue~(\ref{rec2}),
including chiefly the global degree four of the right-hand side,
and the symmetric roles of $Z_k$ and $Z_{n-k}$.
The recursion~(\ref{rec1}) requires only one initial condition, namely $Z_1$.
We set
\beq
Z_1=a.
\label{init}
\eeq
Hereafter the parameter $a$ is chosen to be positive, and viewed as a coupling constant.
The $Z_n$ are then positive, and interpreted as partition functions.

The simplest of all initial conditions, $a=1$,
yields integer values for the $Z_n$:
\beqa
&&Z_1=1,\qquad
Z_2=1,\qquad
Z_3=2,\qquad
Z_4=9,\qquad
Z_5=170,
\nonumber\\
&&Z_6=57\,978,\qquad
Z_7=6\,722\,955\,416,
\eeqa
and so on.
These numbers are listed as sequence number A053294 in the OEIS~\cite{OEIS},
where they are defined as the solution to the very recursion~(\ref{rec1}),
without any motivation nor any useful result.

For a generic initial condition, we have
\beqa
&&Z_2=a^4,
\nonumber\\
&&Z_3=2a^{10},
\nonumber\\
&&Z_4=a^{16}+8a^{22},
\nonumber\\
&&Z_5=8a^{28}+2a^{34}+32a^{40}+128a^{46},
\nonumber\\
&&Z_6=18a^{40}+32a^{46}+128a^{52}+128a^{58}+64a^{64}+1032a^{70}
\nonumber\\
&&{\hskip 12pt}+4352a^{76}+3072a^{82}+16384a^{88}+32768a^{94},
\eeqa
and so on.
The partition function $Z_n$ is a polynomial in $a$ of the form
\beq
Z_n=A_na^{\alpha_n}+\cdots+B_na^{\beta_n}.
\label{hilo}
\eeq
The terms $A_na^{\alpha_n}$ of lowest degrees and $B_na^{\beta_n}$ of largest degrees
will be investigated in detail hereafter
(see~(\ref{bhi}),~(\ref{Bhi}),~(\ref{alres}),~(\ref{aansca})).
Furthermore, the degrees of successive powers of $a$ differ by six, and so
\beq
Z_n=a^{\alpha_n}\,P_n(z),
\eeq
where
\beq
P_n(z)=A_n+\cdots+B_nz^{\Delta_n}
\label{philo}
\eeq
is a polynomial in the variable
\beq
z=a^6
\eeq
with degree
\beq
\Delta_n=\frac{\beta_n-\alpha_n}{6}.
\eeq
We have
\beqa
&&P_1(z)=1,
\nonumber\\
&&P_2(z)=1,
\nonumber\\
&&P_3(z)=2,
\nonumber\\
&&P_4(z)=1+8z,
\nonumber\\
&&P_5(z)=8+2z+32z^2+128z^3,
\nonumber\\
&&P_6(z)=18+32z+128z^2+128z^3+64z^4+1032z^5
\nonumber\\
&&{\hskip 24pt}+4352z^6+3072z^7+16384z^8+32768z^9,
\eeqa
and so on.
Table~\ref{abd} gives the degrees $\alpha_n$, $\beta_n$ and $\Delta_n$ up to $n=14$.

\begin{table}[!ht]
\begin{center}
$$
\begin{array}{|r|r|r|r||r|r|r|r|}
\hline
n & \alpha_n & \beta_n & \Delta_n & n & \alpha_n & \beta_n & \Delta_n \\
\hline
1 & 1 & 1 & 0 & 8 & 64 & 382 & 53 \\
2 & 4 & 4 & 0 & 9 & 88 & 766 & 113 \\
3 & 10 & 10 & 0 & 10 & 112 & 1534 & 237 \\
4 & 16 & 22 & 1 & 11 & 136 & 3070 & 489 \\
5 & 28 & 46 & 3 & 12 & 160 & 6142 & 997 \\
6 & 40 & 94 & 9 & 13 & 184 & 12286 & 2017 \\
7 & 52 & 190 & 23 & 14 & 208 & 24574 & 4061 \\
\hline
\end{array}
$$
\caption{Degrees $\alpha_n$, $\beta_n$ and $\Delta_n$ up to $n=14$.}
\label{abd}
\end{center}
\end{table}

\begin{figure}[!ht]
\begin{center}
\includegraphics[angle=0,width=.7\linewidth,clip=true]{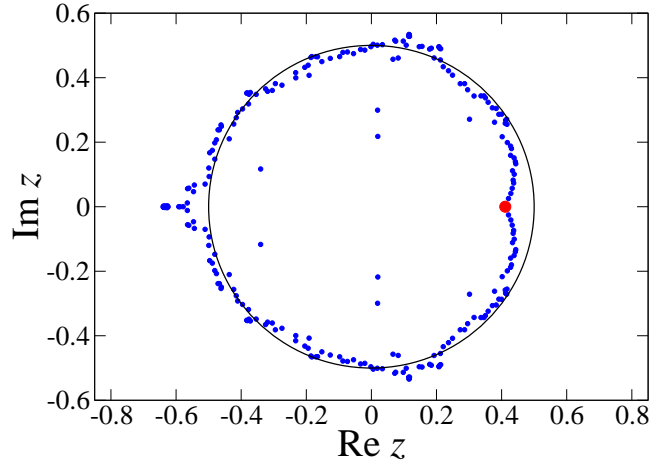}
\caption{\small
Blue symbols: the 237 zeros of the polynomial $P_{10}(z)$ in the complex $z$-plane.
Red symbol: accumulation point of the zeros on the positive real axis at $z_c\approx0.411$.
The black circle has radius $R=1/2$.}
\label{zeros}
\end{center}
\end{figure}

Figure~\ref{zeros} shows a plot of the 237 zeros (or roots) of the polynomial $P_{10}(z)$
in the complex $z$-plane.
Most zeros sit near the circle with radius $R=1/2$, shown in black (see~(\ref{rres})).
The zeros pinch the positive real axis at $z_c\approx0.411$
(red symbol), corresponding to $a_c\approx0.862$.
This observation points toward the emergence of a non-trivial phase diagram
driven by the initial condition~(\ref{init}),
where a critical point at $a_c$ separates a strong-coupling phase ($a>a_c$)
from a weak-coupling one ($a<a_c$).
The detailed analysis performed in Sections~\ref{1dcrit} to~\ref{1dlo} corroborates this picture.

\section{1D model: critical region}
\label{1dcrit}

The 1D model has a critical point at the following value
\beq
a_c=0.862\,322\,847\,096\,001\,235\,198\,066\dots
\label{acres}
\eeq
of the coupling constant.
This critical point is somewhat similar to a separatrix in nonlinear dynamics.
The accuracy of the above number will be commented on at the end of this section.

Right at $a=a_c$, the partition functions should obey a power law of the form
\beq
Z_n\approx C\,n^\gamma.
\eeq
Inserting this asymptotic behavior into~(\ref{rec1}),
and using a continuum setting where the sum over $k$ is replaced by an integral over $y=k/n$,
we successively obtain $\gamma=4\gamma+1$, hence
\beq
\gamma=-\frac{1}{3},
\eeq
and
\beq
C=C^4\int_0^1(y(1-y))^{-2/3}\dd y,
\eeq
hence
\beq
C=\left(\frac{\Gamma(2/3)}{\Gamma^2(1/3)}\right)^{1/3}=0.573\,557\,545\dots
\eeq

Near $a=a_c$,
along the lines of finite-size scaling theory~\cite{cardy,privman},
we expect a scaling behavior of the form
\beq
Z_n\approx C\,n^{-1/3}\bigl(1+D(a-a_c)n^\sigma+\cdots\bigr),
\label{znear}
\eeq
where the crossover exponent
\beq
\sigma=\frac{1}{\nu}
\eeq
is the inverse of the correlation length exponent $\nu$.
Inserting the scaling behavior~(\ref{znear}) into~(\ref{rec1}),
working to first order in $a-a_c$,
and using the same continuum setting as above,
we obtain the following `quantization condition' for the crossover exponent:
\beq
\frac{\Gamma(\sigma+1/3)\Gamma(2/3)}{\Gamma(\sigma+2/3)\Gamma(1/3)}=\frac{1}{4},
\eeq
hence
\beq
\sigma=8.260\,875\,323\dots,\quad\nu=0.121\,052\,547\dots
\eeq
The above scaling analysis yields the exact values of $\gamma$, $C$, $\sigma$ and $\nu$,
but it neither predicts $a_c$ nor the crossover amplitude $D$.

As a consequence of the very large crossover exponent $\sigma$,
the recursion~(\ref{rec1}) is very unstable in the vicinity of $a_c$,
i.e., very sensitive to deviations from $a_c$,
so that the very accurate numerical value of $a_c$ given in~(\ref{acres}) can be obtained.
This level of accuracy will be needed at some places hereafter.

\section{1D model: strong-coupling phase}
\label{1dhi}

In this section we consider the strong-coupling phase $(a>a_c)$ of the 1D model,
which is simpler to analyze than its weak-coupling partner.

The partition functions $Z_n$ grow very fast in this phase,
so that the sum in~(\ref{rec1}) is expected to be
dominated by its extremal terms ($k=1$ and $k=n-1$).
Forgetting about prefactors, this reads
\beq
Z_n\sim Z_{n-1}^2.
\label{hhi}
\eeq
Several heuristic equations of this kind will be met hereafter in the context of 1D models.
The recursion~(\ref{hhi}) does not point toward the existence of oscillations.
It yields an exponentially super-extensive growth law of the form
\beq
\ln Z_n\approx K(a)\,2^n.
\label{hisuper}
\eeq
The quantity $K(a)$ is referred to as the generalized free energy of the model
in its strong-coupling phase.

\subsection{$a\to\infty$ regime}

Let us begin by considering the situation where $a\to\infty$.
In this regime, we have (see~(\ref{hilo}))
\beq
Z_n\approx B_na^{\beta_n}.
\eeq
The above assumption that the sum in~(\ref{rec1}) is dominated by its two extremal terms
is consistent and translates to the recursion relations
\beq
\beta_n=2(\beta_{n-1}+1),\qquad B_n=2B_{n-1}^2\qquad(n\ge3),
\eeq
yielding
\beq
\beta_n=3\,2^{n-1}-2\qquad(n\ge1)
\label{bhi}
\eeq
and
\beq
B_n=2^{2^{n-2}-1}\qquad(n\ge2).
\label{Bhi}
\eeq
These results agree with the form~(\ref{hisuper})
and yield the asymptotic form of $K(a)$ at large $a$, namely
\beq
K_\asy(a)=\frac{3}{2}\ln a+\frac{1}{4}\ln 2.
\label{kas}
\eeq
The first correction term to the above asymptotic form is in $1/a^6$.
The expression~(\ref{kas}) vanishes at
\beq
a_0=2^{-1/6}=0.890\,898\,718\dots,
\label{a0res}
\eeq
whereas $K(a)$ vanishes at $a_c$ (see~(\ref{acres})).
The small difference (3 percent) between~$a_c$ and its strong-coupling approximation~$a_0$
will be demonstrated in Figure~\ref{kplot}.
It can be alternatively illustrated
in terms of the zeros of the polynomials $P_n(z)$, shown in Figure~\ref{zeros} for $n=10$.
The form~(\ref{philo}) implies that the product of all zeros $z_k$ of $P_n(z)$ reads
\beq
\prod_{k=1}^{\Delta_n}z_k=(-1)^{\Delta_n}\frac{A_n}{B_n}.
\eeq
It will be shown below (see~(\ref{alres}) and~(\ref{aansca}))
that $\alpha_n$ and $A_n$ grow much less rapidly than $\beta_n$ and $B_n$.
As a consequence, we have $\Delta_n\approx\beta_n/6\approx2^{n-2}$ and
\beq
R=\lim_{n\to\infty}\left(\frac{A_n}{B_n}\right)^{1/\Delta_n}=\frac{1}{2}.
\label{rres}
\eeq
This number represents the modulus $\abs{z_k}$ of a typical zero of $P_n(z)$ for large $n$.
It is therefore to be expected that most zeros lie near the circle with radius $1/2$.
This prediction is corroborated by Figure~\ref{zeros}.
Finally, $z_0=a_0^6=1/2$ sits right on the above circle,
whereas $z_c=a_c^6$, shown by a red symbol in Figure~\ref{zeros}, is slightly inside that circle.

\subsection{Generic values $(a>a_c)$}

\begin{figure}[!ht]
\begin{center}
\includegraphics[angle=0,width=.6\linewidth,clip=true]{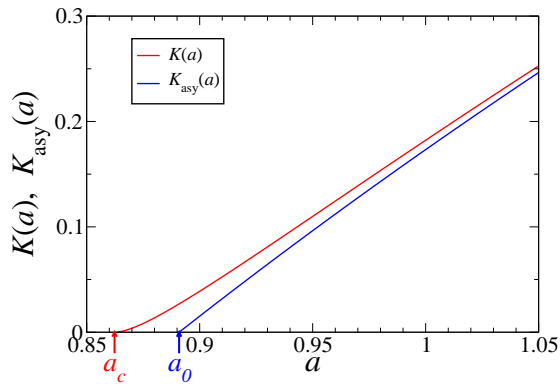}
\caption{\small
Generalized free energy $K(a)$ (red)
and its asymptotic form $K_\asy(a)$ (blue),
plotted against $a$ in some range above the critical point.
Arrows: values $a_c$ and $a_0$ where these functions respectively vanish.}
\label{kplot}
\end{center}
\end{figure}

A numerical iteration of the recursion~(\ref{rec1}) demonstrates that the super-exponential
behavior~(\ref{hisuper}) holds throughout the strong-coupling phase.
Figure~\ref{kplot} shows the generalized free energy $K(a)$ thus obtained
and its asymptotic form~$K_\asy(a)$, plotted against $a$ in some range above the critical point.
The values $a_c$ and $a_0$ where these functions vanish are shown by arrows.
The relatively small difference between these numbers, emphasized just above,
goes hand in hand with the observation that $K_\asy(a)$ provides a good overall representation
of the true $K(a)$, except in the immediate vicinity of the critical point.

\begin{figure}[!ht]
\begin{center}
\includegraphics[angle=0,width=.6\linewidth,clip=true]{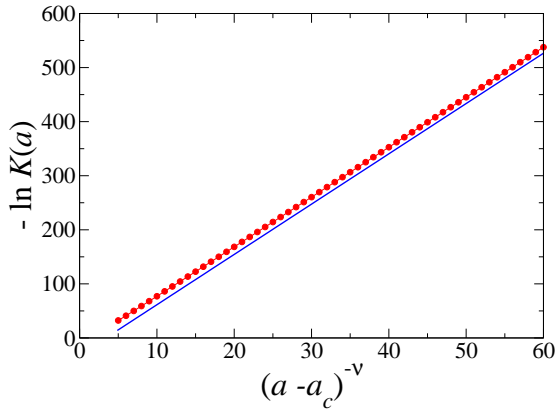}
\caption{\small
Plot of $-\ln K(a)$ against $(a-a_c)^{-\nu}$,
corroborating the critical singularity~(\ref{kess}) of the generalized free energy.
The slope of the blue line yields $C_1\approx9.3$.}
\label{kcrit}
\end{center}
\end{figure}

The critical behavior of the generalized free energy $K(a)$
can be predicted by means of the following scaling argument.
If $a-a_c$ is very small,
the exponential growth~(\ref{hisuper})
only sets in when the sample size $n$ has exceeded a crossover length
diverging as $(a-a_c)^{-\nu}$.
This yields an exponentially small essential critical singularity of the form
\beq
K(a)\sim\exp\left(-C_1(a-a_c)^{-\nu}\right).
\label{kess}
\eeq
This functional form is corroborated by Figure~\ref{kcrit},
showing $-\ln K(a)$ against $(a-a_c)^{-\nu}$.
The slope of the blue line yields
\beq
C_1\approx9.3.
\label{c1res}
\eeq
The rightmost plotted point is the deepest into the critical region,
with $a-a_c\approx2.05\;10^{-15}$ and $K(a)\sim10^{-233}$.
These numbers call for two remarks.
First, only the numerical evaluation of exact equations such as the recursion~(\ref{rec1})
can achieve such accuracy.
Second, it is crucial to use a very accurate value of $a_c$ itself (see~(\ref{acres})).

\section{1D model: weak-coupling phase}
\label{1dlo}

We now turn to the weak-coupling phase of the 1D model $(a<a_c)$.
There, it will be shown that the free energy is extensive, growing as $n^2$,
and exhibits log-periodic oscillations as a function of the sample size $n$,
corresponding to a discrete scaling factor two,
just as the two-dimensional fragmentation model studied in Section~\ref{2d}.

The partition functions $Z_n$ fall off to zero as $n$ increases,
so that the sum in~(\ref{rec1}) is expected to be dominated by its central terms ($k\approx n/2$).
Forgetting about prefactors, this reads
\beq
Z_n\sim Z_{n/2}^4.
\label{hlo}
\eeq
This heuristic relation suggests a growth of $\ln Z_n$ of the form
\beq
\ln Z_n\approx-F(a,x)\,n^2,
\label{lo}
\eeq
where $F(a,x)$ exhibits 1-periodic oscillations
in the logarithmic variable $x$ defined in~(\ref{xdef}).
This quantity is hereafter referred to as the free energy (density) of the model.

\subsection{$a\to0$ regime}

Let us begin by considering the situation where $a\to0$.
In this regime,
the presence of log-periodic oscillations can be explained in simple terms.
We have (see~(\ref{hilo}))
\beq
Z_n\approx A_na^{\alpha_n}.
\label{zlo}
\eeq
We shall successively investigate the degrees $\alpha_n$ and the coefficients $A_n$.

\begin{figure}[!ht]
\begin{center}
\includegraphics[angle=0,width=.6\linewidth,clip=true]{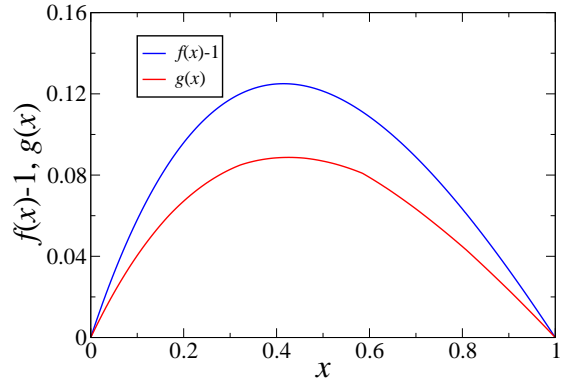}
\caption{\small
The 1-periodic functions $f(x)-1$ (blue) and $g(x)$ (red),
plotted against $x$ over one period.}
\label{fgplot}
\end{center}
\end{figure}

The degrees $\alpha_n$ entering~(\ref{zlo}) obey the recursion
\beq
\alpha_n=2\min_{1\le k\le n-1}(\alpha_k+\alpha_{n-k}).
\label{ainf}
\eeq
These integers are given in Table~\ref{abd} up to $n=12$.
They are listed as sequence number A073121 in the OEIS~\cite{OEIS},
together with useful formulas and references.
In particular, they obey the recursions
\beq
\alpha_{2n}=4\alpha_n,\qquad\alpha_{2n+1}=2(\alpha_n+\alpha_{n+1}).
\eeq
These relations exhibit a discrete scale invariance with scaling factor two.
In particular, $\alpha_n=n^2$ whenever $n=2^m$ is a power of two.
For $n$ in the interval $2^m\le n\le 2^{m+1}$,
the degrees $\alpha_n$ exhibit an exact linear growth in $n$, of the form
\beq
\alpha_n=2^m(3n-2^{m+1}),
\label{alres}
\eeq
so that
\beq
\alpha_n-n^2=(2^{m+1}-n)(n-2^m).
\label{sym1}
\eeq
It is then useful to split the logarithmic variable $x$ defined in~(\ref{xdef})
into its integer and fractional parts according to
\beq
x=\frac{\ln n}{\ln 2}=m+\xi\qquad(0\le\xi\le1),
\eeq
and to introduce another reduced co-ordinate,
\beq
\eta=\frac{n}{2^m}-1=2^\xi-1,
\eeq
that is also in the range $0\le\eta\le1$.
The result~(\ref{alres}) can be exactly recast as
\beq
\alpha_n=n^2 f(x),
\eeq
where $f(x)$ is a 1-periodic function of $x$, given by
\beq
f(x)=2^{-\xi}(3-2^{1-\xi})=1+\frac{\eta(1-\eta)}{(1+\eta)^2}
\label{fxres}
\eeq
in terms of the co-ordinates $\xi$ or $\eta$.
The function $f(x)$ will be plotted in Figure~\ref{fgplot}.
It reaches its minimum $f_\imin=1$ at integer $x$
and its maximum $f_\imax=9/8$ for $\xi=\ln(4/3)/\ln 2$, i.e., $\eta=1/3$.
We have
\beq
f_\ave=\frac{3}{4\ln 2}=1.082\,021\,280\dots,\qquad
f_\osc=\frac{\ln 2}{6}=0.115\,524\,530\dots
\label{fao}
\eeq
We have thus established the validity of the scaling form~(\ref{lo}),
including log-periodic oscillations, to leading order as $a\to0$,
and derived a first estimate of the free energy in the small-$a$ regime,
\beq
F(a,x)\approx f(x)\abs{\ln a}.
\label{f1}
\eeq

\begin{figure}[!ht]
\begin{center}
\includegraphics[angle=0,width=.6\linewidth,clip=true]{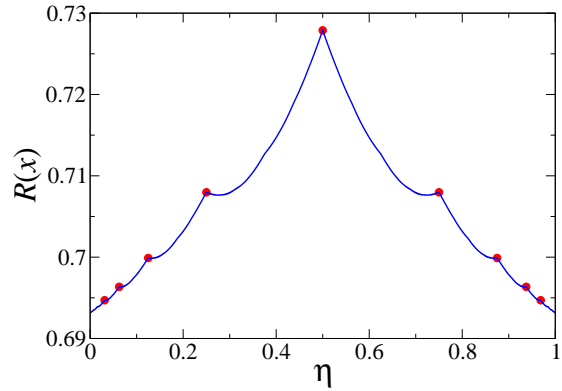}
\caption{\small
Ratio $R(x)$ defined in~(\ref{radef}),
plotted against $\eta$.
Red symbols: main series of cusps at $\eta=2^{-k}$ and $\eta=1-2^{-k}$.}
\label{rfgplot}
\end{center}
\end{figure}

Let us now turn to the analysis of the coefficients $A_n$ entering~(\ref{zlo}).
The minimum in~(\ref{ainf}) turns out to be generically highly degenerate.
More precisely,
for $n$ in the interval $2^m\le n\le 2^{m+1}$,
i.e., $n=2^m+i$ with $0\le i\le 2^m$,
this minimum is reached for all integers $k$ of the form
$k=2^{m-1}+j$, where $j$ runs over the following range:
\beq
\left\{
\matrix{
0\le i\le 2^{m-1}:\hfill & 0\le j\le i,\hfill\cr
2^{m-1}\le i\le 2^m:\quad & i-2^{m-1}\le j\le 2^{m-1}.\hfill
}\right.
\label{kdef}
\eeq
Denoting by $K_n$ the set of integers $k$ defined above, the coefficients $A_n$ obey the recursion
\beq
A_n=\sum_{k\in K_n}A_k^2A_{n-k}^2.
\label{amprec}
\eeq
When $n=2^m$ is a power of two,
the set $K_n$ consists of a single element, $k=n/2$, and so $A_n=1$.
When $n=3\;2^{m-1}$ is half-way between two successive powers of two,
the set $K_n$ is the largest, with $2^{m-1}+1$ elements,
and the coefficient $A_n$ is (locally) maximal.
We have
\beq
A_3=2,\quad
A_6=18,\quad
A_{12}=113\,170,
\eeq
and so on.
More generally, we have the symmetry
\beq
A_n=A_{3\;2^m-n}.
\label{sym2}
\eeq
The coefficients $A_n$ obey the asymptotic growth law
\beq
\ln A_n\approx n^2 g(x),
\label{aansca}
\eeq
where $g(x)$ is a positive 1-periodic function of $x$.
We have thus obtained a more complete estimate of the free energy in the small-$a$ regime:
\beq
F(a,x)\approx f(x)\abs{\ln a}-g(x).
\label{f2}
\eeq
The function $g(x)$ is positive, vanishes at integer~$x$, and its average reads
\beq
g_\ave\approx0.058\,187\,830.
\label{gave}
\eeq
At variance with~(\ref{ainf}),
the recursion~(\ref{amprec}) cannot be solved in closed form,
so that no analytic expression for $g(x)$ is available.
As a consequence of~(\ref{kdef}),
$g(x)$ exhibits cusps at (presumably) all dyadic values $\eta=j/2^k$
of the coordinate $\eta$,
the strongest ones being at $\eta=2^{-k}$ and $\eta=1-2^{-k}$.
Figure~\ref{fgplot} shows the 1-periodic functions $f(x)-1$ and $g(x)$ over one period.
The cusp singularities of $g(x)$ are not visible at this scale.
Figure~\ref{rfgplot} shows the ratio
\beq
R(x)=\frac{g(x)}{f(x)-1}
\label{radef}
\eeq
against $\eta$.
The formulas~(\ref{sym1}) and~(\ref{sym2}) imply that
$R(x)$ is invariant under the change of $\eta$ into $1-\eta$.
The two main series of cusp singularities are now clearly visible (red symbols).

\subsection{Generic values $(a<a_c)$}

A numerical iteration of the recursion~(\ref{rec1}) demonstrates that the scaling
behavior~(\ref{lo}) holds throughout the weak-coupling phase.
We shall focus our attention onto the average $F_\ave(a)$
and the relative magnitude of oscillations~$F_\osc(a)$
of the free energy $F(a,x)$.

In the $a\to0$ regime,
the estimate~(\ref{f2}) shows that the average free energy diverges logarithmically according to
\beq
F_\ave(a)\approx f_\ave\abs{\ln a}-g_\ave,
\label{avezero}
\eeq
where $f_\ave$ and $g_\ave$ are respectively given in~(\ref{fao}) and~(\ref{gave}),
whereas $F_\osc(a)$ goes to the finite limit (see~(\ref{fao}))
\beq
F_\osc(0)=f_\osc=0.115\,524\,530\dots,
\label{osczero}
\eeq
whose value is quite sizeable.

\begin{figure}[!ht]
\begin{center}
\includegraphics[angle=0,width=.6\linewidth,clip=true]{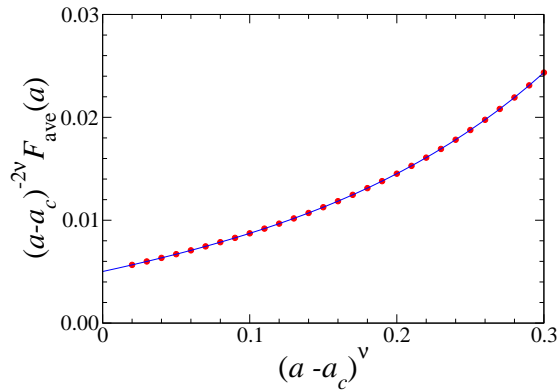}
\caption{\small
Red symbols: $(a_c-a)^{-2\nu}F_\ave(a)$ against $(a_c-a)^\nu$.
Blue curve: polynomial fit yielding $C_2\approx0.0050$.}
\label{fcrit}
\end{center}
\end{figure}

\begin{figure}[!ht]
\begin{center}
\includegraphics[angle=0,width=.6\linewidth,clip=true]{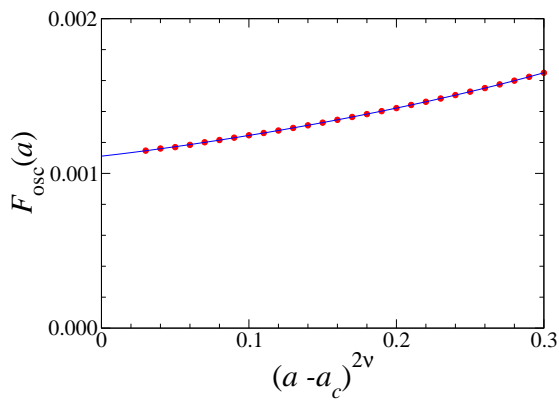}
\caption{\small
Red symbols: $F_\osc(a)$ against $(a_c-a)^{2\nu}$.
Blue curve: polynomial fit yielding $F_\osc(a_c)\approx0.00111$.}
\label{fosccrit}
\end{center}
\end{figure}

In the critical regime $(a\to a_c)$,
scaling theory suggests that the average free energy vanishes according to
\beq
F_\ave(a)\approx C_2(a_c-a)^{2\nu},
\label{avesca}
\eeq
whereas the relative magnitude of oscillations has a finite limit $F_\osc(a_c)$.
These expectations are corroborated by Figures~\ref{fcrit} and~\ref{fosccrit}.
The data plotted there have been extrapolated by means of very accurate 4th-degree polynomial fits.
There is no rationale behind the choice of the abscissas used in these plots:
the chosen powers of $a_c-a$ just turn out to allow for accurate extrapolations.
Figure~\ref{fcrit} shows the combination $(a_c-a)^{-2\nu}F_\ave(a)$ against $(a_c-a)^\nu$,
yielding
\beq
C_2\approx0.0050.
\label{c2res}
\eeq
Comparing this number to the amplitude $C_1$ given in~(\ref{c1res}),
we notice that,
even though $C_1$ is large and $C_2$ is small,
the dimensionless combination $C_1^2C_2\approx0.43$ is of order unity.
Figure~\ref{fosccrit} shows $F_\osc(a)$ against $(a_c-a)^{2\nu}$, yielding
\beq
F_\osc(a_c)\approx0.00111.
\label{oscsca}
\eeq

\begin{figure}
\begin{center}
\includegraphics[angle=0,width=.6\linewidth,clip=true]{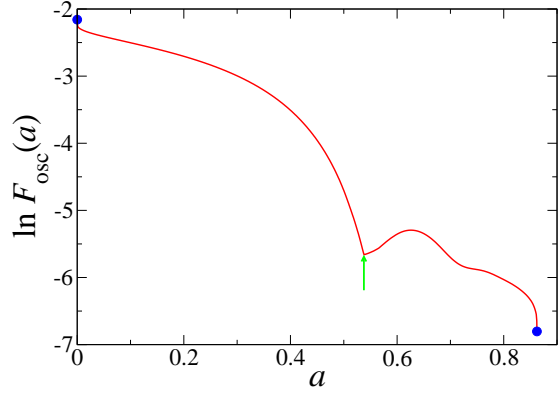}
\caption{\small
Logarithmic plot of $F_\osc(a)$ against $a$ all over the weak-coupling phase.
Blue symbols: limit values~(\ref{osczero}) at $a=0$ and~(\ref{oscsca}) at $a=a_c$.
Green arrow: cusp at $a_*\approx0.537750$.}
\label{fosc}
\end{center}
\end{figure}

As the coupling constant $a$ is increased from 0 to $a_c$,
the average free energy $F_\ave(a)$ decreases steadily from infinity to zero,
interpolating smoothly between the behaviors~(\ref{avezero}) and~(\ref{avesca}).
In the same time, the relative magnitude $F_\osc(a)$ of oscillations
decreases by a factor of order 100 from~(\ref{osczero}) to~(\ref{oscsca}).
Figure~\ref{fosc} shows that $F_\osc(a)$ does not decay monotonically between those two limits.
In particular, it exhibits a cusp at $a_*=0.537\,750\dots$

\begin{figure}
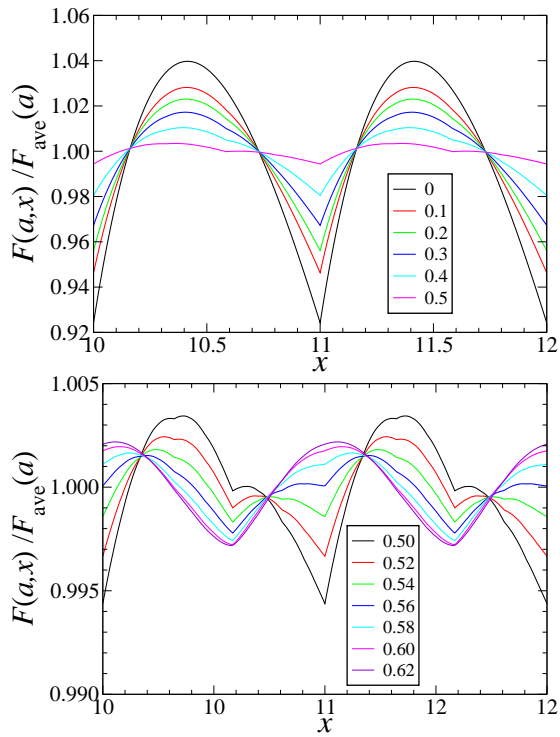

\begin{center}
\includegraphics[angle=0,width=.6\linewidth,clip=true]{funone.eps}

\includegraphics[angle=0,width=.6\linewidth,clip=true]{funtwo.eps}
\caption{\small
Reduced oscillations $F(a,x)/F_\ave(a)$,
plotted against the logarithmic variable $x$ for several $a$ (see legends).
Note the different vertical scales in both panels.}
\label{funs}
\end{center}
\end{figure}

The cusp in Figure~\ref{fosc} has to do with a change in the shape
of the 1-periodic function $F(a,x)$.
Figure~\ref{funs} shows the reduced oscillations $F(a,x)/F_\ave(a)$
against the logarithmic variable $x$.
Two consecutive periods are shown for clarity.
For the smaller values of $a$ (upper panel),
the oscillations strongly decrease in magnitude,
while keeping essentially the shape of $f(x)$ (see~Figure~\ref{fgplot}),
with a cusp around a single minimum at integer $x$.
For~$a$ near~$a_*$ (lower panel),
the structure of the oscillations becomes richer and changes rapidly with $a$.
A second minimum with a cusp develops at $\eta=1/2$, i.e., $\xi=\ln(3/2)/\ln 2=0.584\,962\dots$
This second cusp and the original one have exactly the same height at $a=a_*$.
This degeneracy causes the structure observed in Figure~\ref{fosc}.
For larger $a$ (not shown), the oscillations soon become very small and harmonic.

\newpage

\section{Further 1D examples}
\label{pers}

In this section we propose four other 1D analogues
of the recursion~(\ref{rec1}) for the partition functions of the fragmentation model.
In each case,
we determine whether the model may cross a phase transition when its initial conditions are varied
and whether is exhibits log-periodic oscillations in one of its phases.
These outcomes lead us to introduce, as our fifth example,
what we can think of as the most general recursion of this kind exhibiting log-periodic oscillations.

\subsection*{Example 1}

This first example is obtained by suppressing the squares in the recursion~(\ref{rec1})
defining the 1D model.
We thus obtain the bilinear recursion
\beq
Z_n=\sum_{k=1}^{n-1}Z_kZ_{n-k}\qquad(n\ge2),
\label{e1}
\eeq
with initial condition $Z_1=a$.
In terms of the generating series
\beq
G(z)=\sum_{n\ge1}Z_nz^n,
\label{fdef}
\eeq
this reads
\beq
G(z)=az+G(z)^2,
\eeq
hence
\beq
G(z)=\frac{1-\sqrt{1-4az}}{2}.
\eeq
The partition functions therefore read
\beq
Z_n=C_{n-1}a^n\approx\frac{(4a)^n}{4\sqrt{\pi n^3}},
\label{ze1}
\eeq
where
\beq
C_{n-1}=\frac{(2n-2)!}{n!(n-1)!}
\eeq
is the $(n-1)$-st Catalan number.
These integers have a panoply of combinatorial interpretations.
They are listed as sequence number A000108 in the OEIS~\cite{OEIS},
together with many useful formulas and references.
The exponential growth rate of the $Z_n$ depends smoothly on the initial condition.
This example neither exhibits a phase transition nor oscillations.

\subsection*{Example 2}

This second example is an extension of the previous one,
defined by the trilinear recursion
\beq
Z_n=\sum_{k+l+m=n}Z_kZ_lZ_m\qquad(n\ge3),
\eeq
with initial conditions $Z_1=a$ and $Z_2=b$.
For $a=b=1$, the $Z_n$ are positive integers which count ternary trees.
They are listed as sequence number A019497 in the OEIS~\cite{OEIS}.
In full generality, the corresponding generating series,
defined in analogy with~(\ref{fdef}), obeys
\beq
G(z)=az+bz^2+G(z)^3.
\label{feq}
\eeq
We thus obtain the asymptotic behavior
\beq
Z_n\approx\frac{C}{n^{3/2}\,{z_0}^n},
\eeq
where $z_0$ is the smallest zero of the discriminant of~(\ref{feq}), namely
\beq
\Delta=27z^2(a+bz)^2-4.
\eeq
In the range of physical relevance ($a$ and $b$ positive),
the exponential growth rate $1/z_0$ is positive
and has a smooth dependence on the initial conditions.
This example therefore neither exhibits a phase transition nor oscillations.

\subsection*{Example 3}

The third example is obtained by replacing each square in~(\ref{rec1}) by a higher power, namely
\beq
Z_n=\sum_{k=1}^{n-1}Z_k^pZ_{n-k}^p\qquad(n\ge2),
\label{rec3}
\eeq
with $Z_1=a$.
For any integer $p\ge3$,
the model exhibits the very same phenomenology as the 1D model investigated above,
with a critical point at some $p$-dependent $a_c$.
The strong-coupling phase $(a>a_c)$ is captured by the heuristic equation
\beq
Z_n\sim Z_{n-1}^p,
\label{hhgal}
\eeq
generalizing~(\ref{hhi}) and yielding an exponentially growing free energy of the form
\beq
\ln Z_n\approx K(a)\,p^n.
\eeq
The weak-coupling phase $(a<a_c)$ is captured by the heuristic equation
\beq
Z_n\sim Z_{n/2}^{2p},
\label{logal}
\eeq
generalizing~(\ref{hlo}) and yielding a free energy growing as
\beq
\ln Z_n\approx-F(a,x)\,n^d,
\eeq
where the effective dimension reads
\beq
d=\frac{\ln 2p}{\ln 2},
\eeq
and $F(a,x)$ is a 1-periodic function of the logarithmic variable $x$ defined in~(\ref{xdef}).

\subsection*{Example 4}

The fourth example is obtained by replacing the squares in~(\ref{rec1})
by two different integer powers $p$ and $q$, such that $p>q\ge1$, namely
\beq
Z_n=\sum_{k=1}^{n-1}Z_k^pZ_{n-k}^q\qquad(n\ge2),
\eeq
with $Z_1=a$.
This is the first situation where $Z_k$ and $Z_{n-k}$ play asymmetrical roles.
The model has a critical point at some $a_c$ depending on $p$ and $q$.
The strong-coupling phase $(a>a_c)$ is similar to that of the previous example.
In the weak-coupling phase $(a<a_c)$, the free energy grows as a power of the sample size, namely
\beq
\ln Z_n\approx -F(a)\,n^d,
\label{z4}
\eeq
where the effective dimension $d$ is given by
\beq
\bigl(p^{1/(d-1)}-1\bigr)\bigl(q^{1/(d-1)}-1\bigr)=1.
\eeq
The power-law~(\ref{z4}) is modulated by slowly damped erratic oscillations.
This example therefore exhibits a continuous phase transition,
but no everlasting log-periodic oscillations.

\subsection*{Example 5}

This fifth example is meant to represent the most general recursion
exhibiting log-periodic oscillations,
at least within the class of models under scrutiny.
The study of the previous examples suggests that the following ingredients are necessary:
each partition function enters non-linearly and all parts play symmetric roles.
We are thus led to consider the recursion
\beq
Z_n=\sum_{k_1+\cdots+k_r=n}Z_{k_1}^p\dots Z_{k_r}^p\qquad(n\ge r).
\label{rec5}
\eeq
The recursion~(\ref{rec5}) has to be supplemented with the $r-1$ initial conditions
$Z_n=a_n$ for $n=1,\dots,r-1$.
This model has two integer parameters,
the degree $p\ge2$ and the number $r\ge2$ of parts.
The recursion~(\ref{rec1}) defining the 1D model is recovered for $p=r=2$,
whereas the recursion~(\ref{rec3}) defining Example~3 is recovered for $r=2$ and $p$ arbitrary.

For physically relevant, i.e., positive initial conditions,
the simplest phase diagram one can think of consists of a single critical surface
in the space of initial conditions,
separating a strong-coupling phase from a weak-coupling one.
More complex scenarios allowing for other kinds of behavior in intermediate regimes
are however not entirely ruled out.

The strong-coupling phase is captured by the heuristic equation
\beq
Z_n\sim Z_{n+1-r}^p,
\eeq
generalizing~(\ref{hhi}) and~(\ref{hhgal}),
and yielding an exponentially growing free energy of the form
\beq
\ln Z_n\approx K\,\e^{n\mu},\qquad\mu=\frac{\ln p}{r-1}.
\eeq
The weak-coupling phase is captured by the heuristic equation
\beq
Z_n\sim Z_{n/r}^{pr},
\eeq
generalizing~(\ref{hlo}) and~(\ref{logal}),
and yielding a free energy growing as a power of the sample size
modulated by periodic oscillations, of the form
\beq
\ln Z_n\approx-F(x)\,n^d,
\eeq
where the effective dimension reads
\beq
d=\frac{\ln pr}{\ln r},
\eeq
and $F(x)$ is a 1-periodic function of the logarithmic variable
\beq
x=\frac{\ln n}{\ln r}.
\eeq
The constant $K$ and the function $F(x)$ depend on the $r-1$ initial conditions.

\section{Discussion}
\label{disc}

The stochastic fragmentation model introduced in~\cite{bnk}
sparked our interest in revisiting the subject of log-periodic oscillations.
The fragmentation model is unique in that it combines the following two characteristics.
On the one hand, from a heuristic viewpoint,
it can be expected to exhibit some weak discrete scale invariance,
in analogy with what occurs e.g.~in turbulence or in diffusion-limited aggregation.
On the other hand, the numbers $Z_{m,n}$ of jammed configurations
obey an exact recursion formula.
We have used the latter property to demonstrate by numerical means that the model indeed exhibits
the log-periodic oscillations predicted by the former one.
To our knowledge, this is the first instance of a statistical-mechanical model
where periodic oscillations are reported in the size dependence of a physical quantity.

A 1D analogue of the fragmentation model has then been introduced and investigated in detail.
This 1D toy model has many appealing features.
First of all, it is simple enough to lend itself to a very detailed investigation.
In spite of this, it has a richer behavior than the 2D fragmentation model.
There is a critical value $a_c$ of the initial condition, interpreted as a coupling constant,
that is somewhat similar to a separatrix in nonlinear dynamics.
This critical point separates a strong-coupling phase
where the free energy is super-extensive and does not manifest oscillations,
from a weak-coupling one
where the free energy is extensive and exhibits log-periodic oscillations.
The above characteristics can be established on much firmer ground
than for the 2D fragmentation model
by means of so-called heuristic equations such as~(\ref{hhi}) and~(\ref{hlo}).
Finally, we have generalized the 1D model into a family of models with two integer parameters,
which exhibit essentially the same phenomenology.

This work leaves a number of questions unanswered.
The most pressing one concerns what can be established by either rigorous or analytical means
about log-periodic oscillations in the fragmentation model.
Possible answers range from simply proving their existence
to deriving explicit formulas for the log-periodic functions
that modulate the bulk configurational entropy and related aspect-ratio-dependent quantities.

\section*{Acknowledgments}

It is a pleasure to thank Paul Krapivsky for very stimulating exchanges that motivated this work,
and for having graciously made available to me the graph shown in Figure~\ref{tiling}.
A useful discussion with J\'er\'emie Bouttier and Emmanuel Guitter is also acknowledged.

\bibliography{paper.bib}

\end{document}